\documentclass[journal]{IEEEtran}

\usepackage[T1]{fontenc}
\usepackage{graphicx}
\usepackage{xspace}
\usepackage{cite}

\usepackage{amsmath,amssymb}
\usepackage{algorithm}
\usepackage{algpseudocode}
\usepackage[cmintegrals]{newtxmath}

\usepackage[utf8]{inputenc}
\usepackage{amsmath}
\newtheorem{example}{Example}
\usepackage{pgfplots}
\pgfplotsset{compat=newest}

\usetikzlibrary{calc}
\newlength{\MyFigureWidth}
\newlength{\MyFigureHeight}
\newcommand{\inputtikz}[1]{
			\includegraphics{#1.pdf}
			}
\pgfplotsset{
	every axis legend/.append style={
		legend cell align=left,
		align=left,
		font=\footnotesize
	}
}

\pgfplotsset{
	every axis plot/.append style={
  		line width=1pt,
  		mark size=2pt,
      mark options={solid,line width=0.5pt,fill=white!80!.}
  	}
}

\pgfplotsset{
	every axis/.append style={
		label style={font=\footnotesize},
        tick label style={font=\footnotesize}  
    }
}

\definecolor{StrategyIColor}{rgb}{0.00000,0.44700,0.74100}%
\definecolor{StrategyIIColor}{rgb}{1,0,0}%
\definecolor{InterleaverColor}{rgb}{0,0.4,0}%
\definecolor{MPDMMediumColor}{rgb}{0.929,0.694,0.125}%
\definecolor{MPDMLongColor}{rgb}{1,0,0}%

\pgfplotsset{StrategyI/.style={solid,color=StrategyIColor,mark=diamond*}}
\pgfplotsset{StrategyII/.style={solid,color=StrategyIIColor,mark=pentagon*}}
\pgfplotsset{Capacity/.style={solid,line width=2pt,color=CapacityColor}}
\pgfplotsset{CCDMInf/.style={dotted,color=CCDMInfColor}}
\pgfplotsset{Interleaver/.style={solid,color=InterleaverColor, mark=square*}}
\pgfplotsset{MPDMMedium/.style={solid,color=MPDMMediumColor,mark=pentagon*}}
\pgfplotsset{MPDMLong/.style={solid,color=MPDMLongColor,mark=*}}

\newcommand{\MyVec}[1]{\underline{#1}}

\newcommand{\E}[1]{\ensuremath{\mathbb{E}\left(#1\right)}}
\usepackage{mathtools}
\DeclarePairedDelimiter\abs{\lvert}{\rvert}%
\let\norm\relax
\DeclarePairedDelimiter\norm{\lVert}{\rVert}%
\let\floor\relax
\DeclarePairedDelimiter\floor{\lfloor}{\rfloor}
\makeatletter
\let\oldabs\abs
\def\abs{\@ifstar{\oldabs}{\oldabs*}}
\let\oldnorm\norm
\def\norm{\@ifstar{\oldnorm}{\oldnorm*}}
\let\oldfloor\floor
\def\floor{\@ifstar{\oldfloor}{\oldfloor*}}
\makeatother

\newcommand{\Seqx}{\ensuremath{\MyVec{x}}\xspace}
\newcommand{\Seqy}{\ensuremath{\MyVec{y}}\xspace}
\newcommand{\Seqa}{\ensuremath{\MyVec{a}}\xspace}
\newcommand{\Sequ}{\ensuremath{\MyVec{u}}\xspace}

\newcommand{\nQAMSymsPerFEC}{\ensuremath{l}\xspace}
\newcommand{\runratio}{\ensuremath{\texttt{runratio}}\xspace}
\newcommand{\runratioabs}{\ensuremath{\texttt{runratio}_{||}}\xspace}
\newcommand{\runratioarg}{\ensuremath{\texttt{runratio}_{\measuredangle}}\xspace}

\newcommand{\nsims}{\ensuremath{n_\text{sim}}\xspace}

\newcommand{\nQAM}{\ensuremath{n_\text{QAM}}\xspace}
\newcommand{\KLdiv}{\ensuremath{D_\text{KL}}\xspace}
\newcommand{\kurt}{\ensuremath{\mu_4}\xspace}

\newcommand{\rev}[1]{#1}

\usepackage[hidelinks]{hyperref} 

\begin{document}
\title{Analysis of Nonlinear Fiber Interactions for Finite-Length Constant-Composition Sequences}

\author{Tobias~Fehenberger,~\IEEEmembership{Member,~IEEE},
        David S. Millar,~\IEEEmembership{Member,~IEEE}, Toshiaki Koike-Akino,~\IEEEmembership{Senior Member,~IEEE}, Keisuke Kojima,~\IEEEmembership{Senior Member,~IEEE}, Kieran Parsons,~\IEEEmembership{Senior Member,~IEEE}, and Helmut Griesser,~\IEEEmembership{Member,~IEEE}
        
\thanks{T.~Fehenberger was with Mitsubishi Electric Research Laboratories. He is now with ADVA Optical Networking, Munich, Germany. E-mail: \mbox{tfehenberger@advaoptical.com}.}
\thanks{D.~S.~Millar, T.~Koike-Akino, K.~Kojima and K.~Parsons are with Mitsubishi Electric Research Laboratories. E-mails: millar@merl.com; koike@merl.com; kojima@merl.com; parsons@merl.com.}%
\thanks{H.~Griesser is with ADVA Optical Networking, Munich, Germany. E-mail: \mbox{hgriesser@advaoptical.com}.}
}

\markboth{T.~Fehenberger \MakeLowercase{\textit{et al.}}: Analysis of Nonlinear Fiber Interactions for Finite-Length Constant-Composition Sequences}%
{}

\maketitle

\begin{abstract}
In order to realize probabilistically shaped signaling within the probabilistic amplitude shaping (PAS) framework, a shaping device outputs sequences that follow a certain nonuniform distribution. In case of constant-composition (CC) distribution matching (CCDM), the sequences differ only in the ordering of their constituent symbols, whereas the number of occurrences of each symbol is constant in every output block. Recent results by Amari \textit{et al.} have shown that the CCDM block length can have a considerable impact on the effective signal-to-noise ratio (SNR) after fiber transmission. So far, no explanation for this behavior has been presented. Furthermore, the block-length dependence of the SNR seems not to be fully aligned with previous results in the literature. This paper is devoted to a detailed analysis of the nonlinear fiber interactions for CC sequences. We confirm in fiber simulations the inverse proportionality of SNR with CCDM block length and present two explanations. The first one, which only holds in the short-length regime, is based on how two-dimensional symbols are generated from shaped amplitudes in the PAS framework. The second, more general explanation relates to an induced shuffling within a sequence, or equivalently a limited concentration of identical symbols, that is an inherent property for short CC blocks, yet not necessarily present in case of long blocks. This temporal property results in weaker nonlinear interactions, and thus higher SNR, for short CC sequences. For a typical multi-span fiber setup, the SNR difference is numerically demonstrated to be up to 0.7~dB. Finally, we evaluate a heuristic figure of merit that captures the number of runs of identical symbols in a concatenation of several CC sequences. For moderate block lengths up to approximately 100 symbols, this metric suggests that limiting the number identical-symbol runs can be beneficial for reducing fiber nonlinearities and thus, for increasing SNR.
\end{abstract}

\begin{IEEEkeywords}
Fiber Nonlinearities, Constant-Composition Distribution Matching, Probabilistic Amplitude Shaping, Gaussian Noise Models
\end{IEEEkeywords}

\IEEEpeerreviewmaketitle

\section{Introduction}

\begin{figure}
\begin{center}
\inputtikz{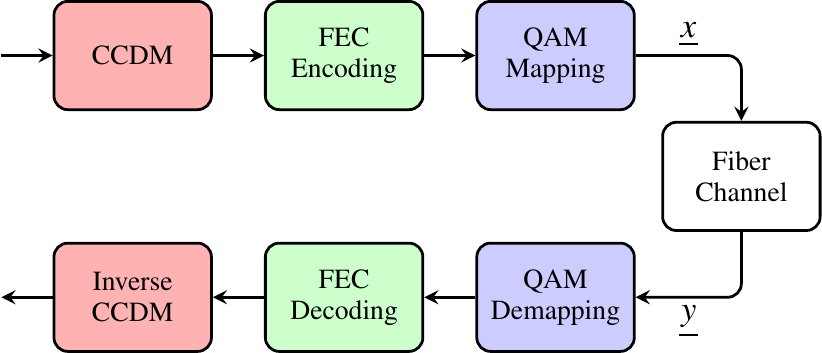}
\caption{\rev{Simplified block diagram of probabilistic amplitude shaping (PAS). This paper studies the impact of finite-length sequences $\Seqx$, which are generated from the amplitude sequences output by the CCDM, on the fiber output $\Seqy$, specifically on the effective SNR after fiber transmission. A complete block diagram for PAS can be found in \cite[Fig.~3]{Boecherer2015TransComm_ProbShaping}}.}
\label{fig:block_diagram}
\end{center}
\end{figure}

Probabilistic amplitude shaping (PAS) \cite{Boecherer2015TransComm_ProbShaping} has become the de-facto standard coded modulation (CM) framework for realizing probabilistic shaping, offering about 1 dB of signal-to-noise ratio (SNR) improvement and rate adaptivity without major modifications to an existing CM architecture that was designed for unshaped, i.e., uniform, signaling. An important part of PAS is the shaper that transforms the uniform data sequence to shaped amplitudes at the transmitter. The inverse operation at the receiver recovers the initially sent data from the shaped amplitudes, assuming that all transmission errors have been corrected by forward error correction (FEC), see Fig.~\ref{fig:block_diagram} for a block diagram. The most prominent shaping system is constant-composition distribution matching (CCDM) \cite{Schulte2016TransIT_DistributionMatcher}, which is used in the initial PAS proposal \cite[Sec.~V]{Boecherer2015TransComm_ProbShaping}. As the name suggests, a CCDM of a certain length outputs sequences of constant composition (CC), i.e., which have a constant number of occurrences of each amplitude and are hence permuted version of each other. More advanced distribution matchers include multi-set partition distribution matching (MPDM) \cite{Fehenberger2019TCOM_MPDM} and multi-composition codes \cite{Pikus2019Arxiv_MCBLDM}. Other fixed-length shapers that are not directly related to the concept of compositions include enumerative sphere shaping (ESS) \cite{Gultekin2019Arxiv_ESS} and shell mapping \cite{laroia1994}, both of which establishing an energy-efficient indexing of all sequences within a sphere of fixed radius, \rev{as well as prefix-free code distribution matching (PCDM) with framing \cite{choPrefixFreeCodeDistribution2019}}.

Probabilistically shaped quadrature amplitude modulation (QAM) is well-studied for the nonlinear fiber channel in theory \cite{Dar2014ISIT_nonlinearShaping}, transmission simulations \cite{Fehenberger2016JLT_ShapingQAM,sillekensSimpleNonlinearityTailoredProbabilistic2018,amariIntroducingEnumerativeSphere2019,Cho2016ECOC_ShapingNonlinearTolerance} and experiments \cite{Buchali2016JLT_ProbShapingExp,Cho2016ECOC_ShapingNonlinearTolerance,ghazisaeidiAdvancedLBandTransoceanic2017,maherConstellationShaped662017}. Temporal shaping of quaternary phase shift keying (QPSK) over a few time slots has been performed in \cite{yankovTemporalProbabilisticShaping2017}.
In \cite{amariIntroducingEnumerativeSphere2019}, which is the motivation for this manuscript, the authors investigate ESS and CCDM in multi-span wavelength-division multiplexing (WDM) simulations and find that the effective SNR after fiber transmission is inversely proportional to the block lengths of the employed CCDM and ESS. \rev{Very recently, it has been experimentally confirmed that short CCDM blocks indeed give higher SNR than long sequences \cite[Fig.~3b]{goossensFirstExperimentalDemonstration2019}.} Unfortunately, no reasons for this relation are presented in these papers or, to the best of our knowledge, anywhere else. Upon re-analysis of the simulation data of our previous work \cite{Fehenberger2019OFC_MPDM} where we focused on the performance of shaped signaling with low-density parity-check FEC, we are able to confirm also in that paper that the average SNR varies with CCDM block length, which had not been mentioned therein. This finding of block-length dependent SNR poses questions on previously reported results, and also some challenges for future fiber models on which we will comment on briefly.

Regarding the latter, a block-length dependence of SNR after fiber transmission is not covered by the prominent enhanced Gaussian noise (EGN) model \cite{Carena2014OptExp_EGNmodel,Dar2013OptExp_PropertiesNLIN} which considers only time-averaged statistics by assuming infinite-length blocks. The impact of the channel input on the fiber nonlinearities is described by the fourth and sixth moment of the QAM constellation, but finite-length temporal properties are not included. The time-domain model presented in \cite{Agrell2014JLT_FiniteMemory} incorporates temporal effects, i.e., memory, by including the power of past and future symbols, yet not their phase information. The models proposed in \cite[Sec.~II]{Dar2013OptExp_PropertiesNLIN}, \cite{Dar2016JLT_PulseCollision}, \cite{golaniModelingBitErrorRatePerformance2016} include all information of previous and past symbols. All of these models are arguably prohibitively complex to evaluate for many practical setups in which long memory is to be captured. Powerful models that are relatively simple to evaluate could thus be helpful in better understanding time-dependent nonlinear interactions, and in predicting the system performance particularly when short-length shaping is applied. We will set this open research problem aside and focus on understanding the origin of the SNR dependence on block length in the following.
    
Previous fiber simulations \cite{Fehenberger2016JLT_ShapingQAM,Fehenberger2017OFC_ShapingSNRAIR}, \cite[Fig.~2]{kashiExtendingNonlinearSNR2019} and experiments \cite[Fig.~8]{ghazisaeidiAdvancedLBandTransoceanic2017}, \cite{Renner2017JLT_ShortReachShaping} found that the average SNR after transmission over the nonlinear fiber channel is smaller for shaped QAM than for uniform QAM. \rev{None of these papers, however, specifies the length of the employed DM, which suggests that either very long blocks were used where rate loss is negligibly small, or that a DM was emulated by drawing the QAM symbols randomly according to the target distribution. With the simulations results presented in \cite{amariIntroducingEnumerativeSphere2019} and the detailed analysis presented herein, the general finding that the SNR of shaped QAM is worse than that of uniform QAM must be further confined as this appears to be only true when sufficiently long shaping blocks are used. This manuscript is devoted to a detailed study into the origin of this block-length dependence.}

We first analyze the generation of QAM symbols from CC amplitude sequences, as it is carried out for CCDM in the PAS framework. The main part of this work is then devoted to a study of temporal properties of a concatenation of CC sequences for varying DM block lengths. After a theoretical assessment, we perform extensive fiber simulations in order to analyze the dependence of the effective SNR after nonlinear fiber transmission, which we simply refer to as SNR, on the block length of these CC sequences. Throughout this work, we focus on CCDM of varying length but with a constant per-sequence distribution as to isolate the sequence properties (and thereby the nonlinear effects) that are present for different finite block-lengths. The study of \rev{low-rate-loss} shapers, such as ESS,  MPDM \rev{or PCDM}, is more complex since their outputs \rev{in general} do not fulfill the CC property and is left for future work. The considered setting of having a fixed distribution at varying block length differs from \cite{Fehenberger2019ECOC_NLI_MPDM} where the inverse case of SNR fluctuations for varying per-block composition at a fixed block length is considered.

The findings in this paper are as follows.
We first demonstrate that the distribution of the QAM symbols, of each which is obtained from combining two amplitudes, can differ from the infinite-length expectation for very short block lengths. This is a direct result of how the one-dimensional shaped ASK signals are paired to QAM symbols in the PAS framework, and the first reason for an increased effective SNR at ultra-short blocks.
The main reason for the SNR variations, however, is shown to be due to intra- and inter-block nonlinear effects, i.e., temporal nonlinear interactions within a shaping block and over adjacent blocks. By introducing an interleaver for each FEC codeword, we demonstrate that the SNR-dependence of the block length disappears almost entirely. When comparing short and long block lengths without interleaving, it becomes evident that some kind of temporally structured shuffling of the shaped amplitudes, which is implicitly generated with short-length CCDM, leads to reduced nonlinear effects and thus higher SNR than for long block-lengths. Using a heuristic metric that represents the number of identical-symbol runs in a compound sequence, we present strong arguments that, up to moderate block lengths, it is mainly the limited number of runs of identical symbols that gives the SNR improvement. To the best of our knowledge, this manuscript is the first detailed study of the characteristic effects that temporal structures in the transmit sequences have on the generation of nonlinear fiber interference.

\section{QAM Symbols Generated From Constant-Composition Amplitude Sequences}

\subsection{Fundamentals of Probabilistic Amplitude Shaping}
In PAS, the shaped amplitude output of the CCDM is combined with uniform bits, which are the FEC parity and possibly some data bits, into QAM symbols, see Fig.~\ref{fig:block_diagram} for a coarse block diagram. More details on PAS can be found in \cite[Sec.~IV]{Boecherer2015TransComm_ProbShaping}. The output of a CCDM is fully described by its integer-valued composition, which describes exactly how often an amplitude occurs in every block. All possible CCDM output sequences for a fixed composition are permutations of each other, with the number of permutations given by the multinomial coefficient.
\begin{example}[CC Sequences]
The amplitude distribution $[0.4, 0.3, 0.2, 0.1]$ for the CCDM block length $n=10$ gives the composition $[4,3,2,1]$, meaning that the first amplitude occurs exactly four times in the output sequence of length 10, the second symbol three times and so forth. \rev{There are $\binom{10}{4}\cdot\binom{6}{3}\cdot\binom{3}{2}\cdot\binom{1}{1}\cdot= 12600$ different sequences with this composition.} Increasing the block length to $n=100$ gives the composition $[40,30,20,10]$ and $\sim 4.9\cdot10^{52}$ different sequences.
\end{example}

\begin{figure}[t]
\begin{center}
\inputtikz{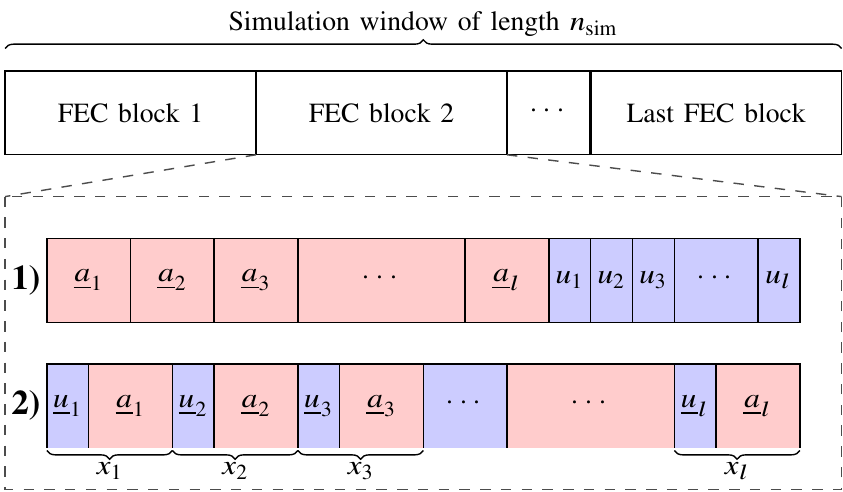}
\end{center}
\caption{Temporal structure of a typical simulation in the PAS framework: one simulation windows consists of several FEC blocks, which in turn contain one or more DM blocks. The block structures \textbf{1)} and \textbf{2)} are equivalent representations of the constituent symbols of a FEC word. In \textbf{1)}, the parity of the systematic encoder is at the end whereas \textbf{2)} has the parity information distributed over the codeword as prefix $\MyVec{u}_i$ of the shaped amplitudes $\MyVec{a}_1$, which is how QAM symbols $\MyVec{x}_1$ are generated in PAS.}
\label{fig:block_structure}
\end{figure}

In Fig.~\ref{fig:block_structure}, the nesting of several DM codewords in one FEC block is shown. The overall considered simulation run (or experimental batch) contains several of these FEC blocks. The two rows \textbf{1)} and \textbf{2)} illustrate how the uniform sign bits $\Sequ$ and the shaped amplitudes $\Seqa$ are grouped within a FEC word.\footnote{Notation: Sequences are underlined with their elements in regular font, i.e., $\Seqx=[x_1,\dots,x_n]$. A discrete random variable $X$ has the probability mass function (PMF) $P_X$ on $\mathcal{X}$.} The canonical view of a systematic code is \textbf{1)} where the parity bits go at the end of each FEC block. Row \textbf{2)} of Fig.~\ref{fig:block_structure} is closer to the structure that is actually transmitted over the channel: In PAS with QAM signaling, two uniform bits $\MyVec{u}_i$ that determine the quadrant are prefix to a pair of shaped amplitudes $\MyVec{a}_i$, which represents the shaped QAM payload, and combining these gives a shaped QAM symbol $\MyVec{x}_i$. It is important to realize that the generated QAM sequences is thus not CC (although the amplitudes are) because the quadrant identifiers are only approximately uniformly distributed. We further note that a pair of shaped amplitudes is required for each QAM symbol. These can be combined in two different ways, as we will discuss next.

\subsection{Two Strategies for Pairing Shaped Amplitudes}\label{ssec:pairings}
In the following, we study the QAM distributions that are obtained by combining the output amplitudes of varying-length CCDM. Intuitively, one would expect that the empirical distribution of the QAM symbols always follow the Cartesian product of the two-sided shaped amplitude shift keying (ASK) distributions. Firstly, note that this expectation only holds on average because the quadrant identifier, i.e., the first two bits of every QAM symbols as shown in Fig.~\ref{fig:block_structure} row \textbf{2)}, is in general random data.\footnote{In PAS, the quadrant identifier comes at first from the redundancy bits of the FEC codeword. If the FEC rate is chosen such that not enough parity bits are generated, uniformly distributed data bits are used in addition, see \cite[Sec.~IV-D]{Boecherer2015TransComm_ProbShaping}.} This shows that the distribution of the QAM symbols is \emph{not} constant-composition, even if the 1D amplitudes are generated with a CCDM.

\begin{figure}[t]
\begin{center}
\inputtikz{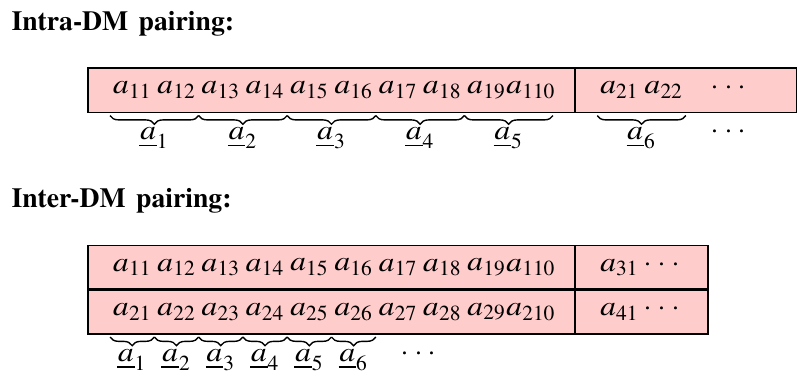}
\end{center}
\caption{Two strategies for combining two shaped amplitudes into the payload of a QAM symbol, illustrated for $n=10$. Top: Intra-DM pairing where two adjacent amplitudes are combined within a DM block. Bottom: Inter-DM pairing where the two amplitudes to be paired come from different DM outputs.}
\label{fig:example_PAS_pairings}
\end{figure}

We now set this aspect of random quadrant identifiers aside as it will average out over time, and analyze constraints in the 2D distributions that stem from short-length CCDM. First, we investigate in detail how the QAM payload, i.e., the shaped data within each 2D symbol, is generated. In the PAS framework, as explained above, two ASK symbols are combined with a two-bit prefix, which determines the quadrant, into a QAM symbol. This means that the DMs must output twice as many shaped amplitudes as QAM symbols are transmitted. Note that this holds not only for CCDM but in general for any PAS shaper that generates amplitudes. An important aspect is how the amplitude symbols are paired into the QAM payload, which has, to the best our knowledge, not been studied yet. Figure~\ref{fig:example_PAS_pairings} shows two different strategies. In the first approach, illustrated at the top of Fig.~\ref{fig:example_PAS_pairings} and denoted as \emph{intra-DM pairing}, the amplitudes sequences are simply concatenated and each QAM payload $\Seqa$ is generated by combining consecutive symbols within a DM output. The pairing offset is fixed to zero, i.e., the first symbol is paired with the second symbol, the third with the fourth and so forth. A different strategy, referred to as \emph{inter-DM pairing}, is shown at the bottom of Fig.~\ref{fig:example_PAS_pairings}. Two DM outputs are stacked on top of each other and two amplitudes, one from each DM output, are combined. Hence, the QAM payload $\Seqa$ for this pairing strategy is the combination of two DM outputs that are independent in the sense that the CC constraint is fulfilled by each of them individually, yet not jointly. Note that the original PAS proposal does not specify which strategy to use but only states that QAM constellations can be generated "by mapping two real ASK symbols to one complex QAM symbol" \cite[p.~4652]{Boecherer2015TransComm_ProbShaping}. We will illustrate in the following example that for intra-DM pairing, a different QAM distribution than expected can be obtained, in particular for very short blocks.

\begin{figure}
\begin{center}
\inputtikz{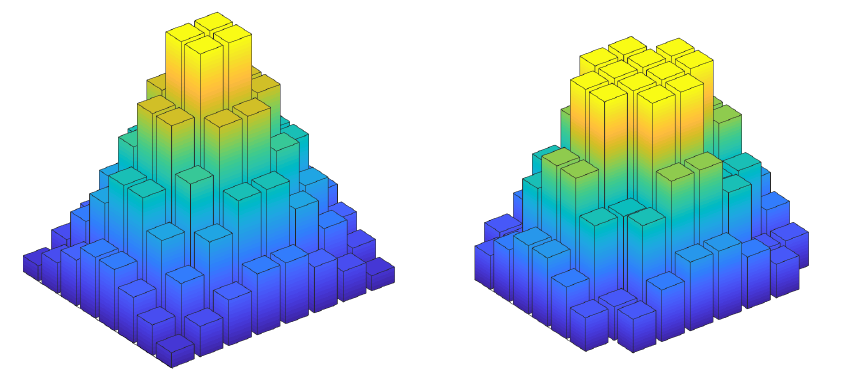}
\end{center}
\caption{64QAM (left) expected for infinite-length sequences vs. 60QAM (right) obtained for the amplitude distribution $[0.4, 0.3, 0.2, 0.1]$ and $n=10$, see Example~\ref{ex:qam60} for details.}
\label{fig:Constellations_64QAM_60QAM}
\end{figure}

\begin{example}[60QAM instead of 64QAM]\label{ex:qam60}
We consider the amplitude distribution $[0.4, 0.3, 0.2, 0.1]$ and $n=10$, which gives the composition $[4,3,2,1]$. The expected probabilities of the shaped 64QAM, which are computed from the two-sided amplitude distribution as Cartesian products, are
\begin{equation*}
\begin{bmatrix} 
 0.010 & 0.020 & 0.030 & 0.040 \\
 0.020 & 0.040 & 0.060 & 0.080 \\
 0.030 & 0.060 & 0.090 & 0.120 \\
 0.040 & 0.080 & 0.120 & 0.160
\end{bmatrix},
\end{equation*}
shown and normalized for the 16 constellation points in the top-left quadrant. With intra-DM pairing, which we recall to be the pairing of adjacent symbols within a DM output, the outermost QAM symbol is only obtained when the highest-energy amplitude occurs as two consecutive symbols within one CC sequence. In this short-length example, this is impossible to happen because the highest-energy sequence occurs only once per DM output. Therefore, the outermost QAM symbols have zero probability of occurrence instead of the expected 0.01. Rather than the expected 64QAM, a 60QAM distribution is generated with the per-quadrant PMF 
\begin{equation*}
\begin{bmatrix} 
0    & 0.022 & 0.034 & 0.045 \\
0.022 & 0.022 & 0.067 & 0.090 \\
0.034 & 0.067 & 0.067 & 0.133 \\
0.045 & 0.090 & 0.133 & 0.133 
\end{bmatrix}.
\end{equation*}
The full PMFs of the expected 64QAM and the obtained 60QAM are shown in Fig.~\ref{fig:Constellations_64QAM_60QAM}. As noted above, the outer-most, i.e., highest-energy QAM symbol, cannot occur at all, while some of inner QAM points are more likely to appear than expected. 
\end{example}

\subsection{Kullback-Leibler Divergence}
\begin{figure}[t]
\begin{center}
\inputtikz{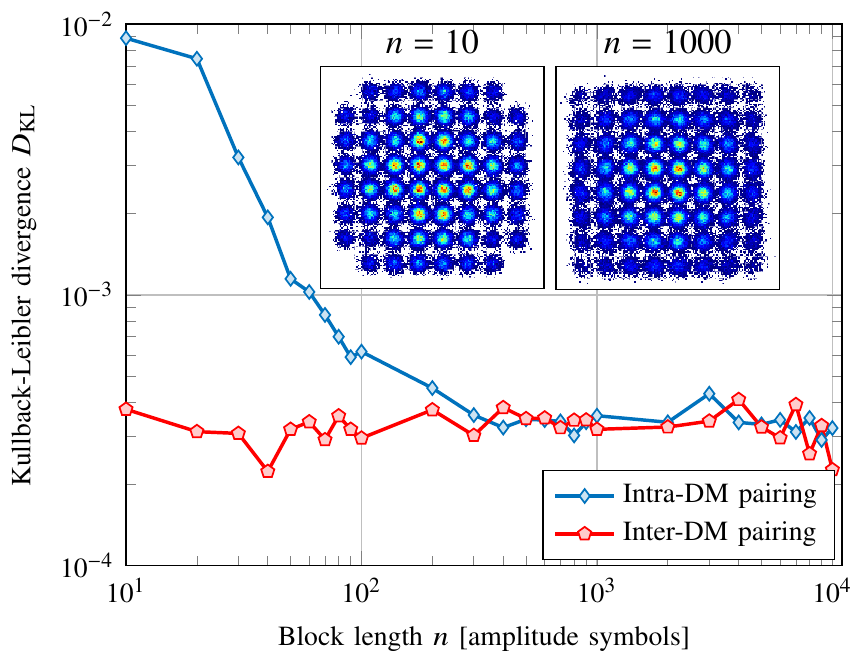}
\end{center}
\caption{KL divergence between the expected QAM PMF and the PMF that is obtained from constant-composition amplitude sequences, as a function of the block length $n$. Higher KL divergence means larger difference in distributions. Insets: Received constellations at approximately 19 dB SNR for the intra-DM amplitude pairing and two different block lengths. The top left-most marker and the left inset correspond to Example~\ref{ex:qam60}.}
\label{fig:CCDM_KLdiv_vs_N}
\end{figure}

The above Example~\ref{ex:qam60} is an extreme case of ultra-short block length which leads to certain QAM symbols not occurring at all. In Fig.~\ref{fig:CCDM_KLdiv_vs_N}, we generalize this analysis to a wider range of block lengths and to both  intra-DM and inter-DM pairing that were introduced in the context of Fig.~\ref{fig:example_PAS_pairings}. The Kullback-Leibler (KL) divergence between the expected 2D QAM distribution and the actually obtained PMF is used as measure of difference in distributions \cite[Sec.~2.3]{CoverThomas2006Book_IT}. We generated $\nQAM=100000$ QAM symbols $x_i$ for $i = \{1,\dots,\nQAM\}$ from CC sequences of different length $n$ and computed the KL divergence \KLdiv as an average over all symbols as

\begin{equation}\label{eq:KLdiv}
\KLdiv = \sum_{x' \in \mathcal{X}} P_X(x') \log_2 \frac{P_X(x')}{\frac{1}{\nQAM}\sum_{i=1}^{nQAM} \delta(x_i,x')},
\end{equation}
where $P_X(\cdot)$ is the expected QAM distribution obtained from the Cartesian product of the shaped ASK PMF, and the denominator is the empirical distribution obtained over $\nQAM$ symbols. The Kronecker delta function is defined as
\begin{equation}\label{eq:deltafun}
\delta ({a,b}) = \begin{cases}
0 &\text{if } a \neq b,   \\
1 &\text{if } a=b.   \end{cases}
\end{equation}
For the results of Fig.~\ref{fig:CCDM_KLdiv_vs_N}, the amplitude distribution was fixed to $[0.4, 0.3, 0.2, 0.1]$, corresponding to the QAM PMF used in Example~\ref{ex:qam60}. Using intra-DM amplitude-pairing, a clear difference between the expected and the empirical QAM distribution is observed. When the block length is increased to a few hundred symbols, the KL divergence saturates at approximately $4\times10^{-4}$. The reason for it not reaching zero is the randomness in the QAM quadrant bits that leads to non-CC QAM sequences although the 1D amplitudes are indeed CC. Inter-DM pairing, which combines the outputs of two independent DMs, gives approximately constant KL divergence and most notably, does not lead to the large KL divergence at short block lengths. This can be explained by the fact that independent DM outputs are combined and thus, no constraints are imposed on which amplitude pairs can be generated. 

\subsection{Interpretation within the EGN Model}
\begin{figure}
\begin{center}
\inputtikz{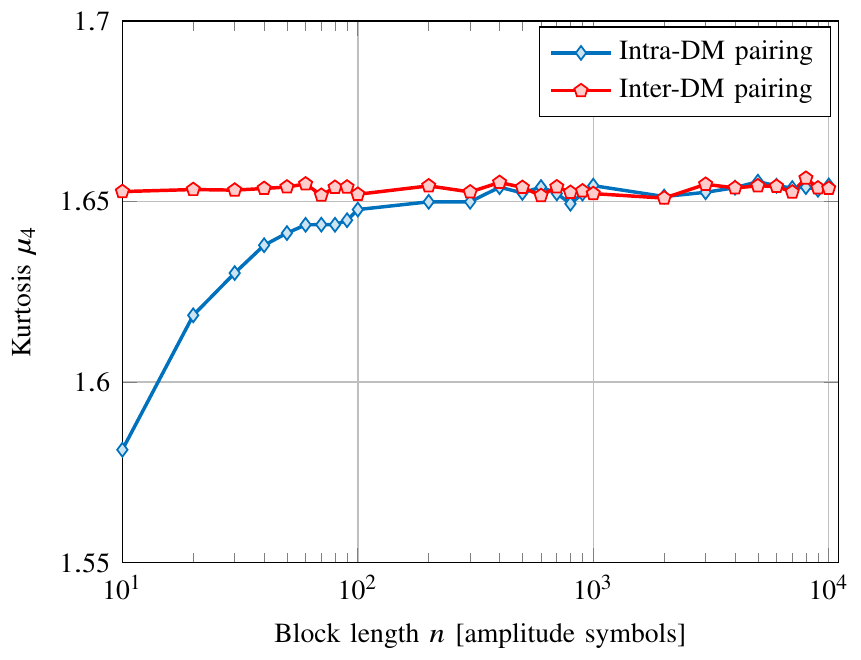}
\end{center}
\caption{2D kurtosis versus CCDM block length for the two pairing strategies introduced in Fig.~\ref{fig:example_PAS_pairings}. The considered amplitude distribution is $[0.4, 0.3, 0.2, 0.1]$, which gives shaped 64QAM.}
\label{fig:CCDM_Kurtosis_vs_N}
\end{figure}

This finding of infeasible and rare QAM symbols is now applied to fiber transmission. The EGN model predicts an SNR decrease with increase in kurtosis \cite{Fehenberger2016JLT_ShapingQAM}, which is the standardized fourth moment of the QAM input $X$ and is defined as
\begin{equation}
\kurt = \frac{\E{|X-\mu|^4}}{\E{|X-\mu|^2}^2},
\end{equation}
where $X$ has mean $\mu$ and $\E{\cdot}$ denotes expected value. Figure~\ref{fig:CCDM_Kurtosis_vs_N} shows the kurtosis \kurt over block length $n$ for the above setting where KL divergence was analyzed. We observe that in the length regime of up to approximately $n=100$ symbols where intra-DM pairing caused a large KL divergence (see Fig.~\ref{fig:CCDM_KLdiv_vs_N}), the kurtosis is reduced, whereas it is constant around 1.65 for inter-DM pairing. Hence, some of the SNR improvement for short block lengths that will be presented in Sec.~\ref{sec:num_results} can be attributed to the empirically generated QAM distribution having smaller kurtosis than for long blocks. However, as we will see in Sec.~\ref{sec:num_analysis}, this only partly explains the SNR dependence on shaping block length that extends into length regimes where the kurtosis has reached its asymptotic value.

\section{Properties of Constant-Composition Sequences for Varying Block Lengths}
In the following, we analyze some important properties of compound CC sequences that are generated by varying the constituent CCDM length while keeping the per-block distribution fixed. This setting allows to isolate some properties that will be relevant for the study of the nonlinear fiber channel in Sec.~\ref{sec:num_analysis}.

\subsection{Inherent Shuffling}\label{ssec:diff_sequences}
An important realization when comparing block lengths for the same distribution is that all sequences that are generated by concatenating several short CCDM outputs can also come from a single long DM. The inverse statement, however, is not true as there are sequences of a long DM that cannot be the result of several short DMs. When looking at the temporal properties of these two cases, an obvious difference is that for many short CCDMs, long runs of consecutive symbols are prohibited since the CC principle must be fulfilled within every constituent sequence. This implies that using multiple short-DM blocks imposes some kind of shuffling of the amplitudes and simultaneously prevent a concentration of identical symbols, which does not necessarily have to occur for long blocks. We illustrate this property in the following example.

\begin{example}[Limited Concentration of Identical Symbols]\label{ex:sequences}
Consider two CCDMs with $n=10$ and $n=1000$, both targeting the distribution $[0.4, 0.3, 0.2, 0.1]$ for the amplitudes $[\alpha, \beta, \gamma, \delta]$. The overall length of the compound sequence to be generated is set to 1000~symbols, which means that 100 CCDM outputs are concatenated for $n=10$, whereas for the long DM a single output is sufficient. Considering the short-DM case, the rarest amplitude $\delta$ must occur exactly once in every ten-symbol block. Over the entire sequence of length 1000, this gives a maximum temporal spacing of two occurrences of $\delta$ of 19 when one CCDM output begins with $\delta$ and the next one ends with this symbol. For the most likely symbol $\alpha$, at most eight consecutive occurrences are feasible, which would be the case if a DM output ends with four times $\alpha$ and the next one begins with four times $\alpha$. This shuffling, or equivalently limited concentration of symbols, holds over all 100 CCDM blocks and thus over the entire sequence of length 1000 as long as no interleaving is performed. In contrast, much longer runs of the same amplitude are feasible for the single CCDM output with $n=1000$, and rare symbols such as $\delta$ can be spaced further apart. An extreme case of a compound sequence that is impossible to be generated by several short DMs is when all amplitudes are lumped together, i.e., the output is $\Seqa{}=[\underbrace{\alpha,\dots,\alpha}_{400\times},\underbrace{\beta,\dots,\beta}_{300\times},\underbrace{\gamma,\dots,\gamma}_{200\times},\underbrace{\delta,\dots,\delta}_{100\times}]$. 
\end{example}

We emphasize that these temporal properties are the only difference between using many short DMs and a few (or just one) long DM when the same distribution and the same compound sequence length is used. Thus, the observation that the effective SNR after fiber transmission varies with CCDM block length (while keeping all other parameters unchanged) can only be attributed to these time-dependent properties. Further note that, for a purely linear channel with additive white Gaussian noise only, such length aspects would not have any impact on the channel and thus, the SNR measured after transmission would remain fixed.


\subsection{Run Ratio}
In the following, we introduce a heuristic for compound CC sequences that is intended to describe, to some extend, these time-dependent properties. The idea is that the shuffling of a shaped CC sequence can be partially represented by how often a transmit symbol differs from its precursor, or equivalently by the number of \emph{runs} within that sequence. The concept of runs is known from run tests used for determining randomness, see for example \cite[Ch.~12]{bradley1968distribution} and \cite{waldTestWhetherTwo1940}. We normalize the number of runs by the number of simulated QAM symbols \nsims in the overall sequence in order to obtain a figure of merit that allows to compare different compound sequence lengths with respect to their concentration of symbols. We refer to this metric as \emph{run ratio} and define it as
\begin{equation}\label{eq:runratio}
\runratio = \frac{1}{\nsims} \left( 1+ \sum_{i=2}^{\nsims} \bar{\delta} \left( x_{i-1},x_{i} \right) \right),
\end{equation}
where $\bar{\delta}(a,b)$ is the inverse of the Kronecker delta function of \eqref{eq:deltafun}, 
\begin{equation}\label{eq:deltafuninverse}
\bar{\delta}({a,b}) = \begin{cases}
1 &\text{if } a \neq b,   \\
0 &\text{if } a=b.   \end{cases}
\end{equation}
The factor $\frac{1}{\nsims}$ in \eqref{eq:runratio} is the aforementioned normalization and the addend 1 in front of the sum ensures that the run beginning with the first symbol $x_1$ is also counted. A large \runratio means that there are few consecutive occurrences of identical symbols, whereas a small \runratio indicates that a sequence has many runs of identical symbols and thus, little shuffling.
\begin{example}[Run Ratios]\label{ex:runratios}
Consider a sequence of $\nsims=10$ QPSK symbols with the four unit-energy constellation points denoted as $[\alpha,\beta,\gamma,\delta]$. The sequences $\Seqa_1=[\alpha \beta \gamma \delta \alpha \beta \gamma \alpha \beta \alpha ]$ and $\Seqa_2=[\alpha \beta \alpha \gamma \alpha \delta \gamma \beta \alpha \beta ]$ each have 10 runs (of length 1), and thus a run ratio of $\runratio=10/10=1$. The sequence $\Seqa_3=[\alpha \alpha \alpha \alpha \beta \beta \beta \gamma \gamma \delta ]$ has 4 runs, giving a run ratio of $4/10=0.4$.
\end{example}

We further introduce an absolute-value version of the run ratio of \eqref{eq:runratio} and define it as
\begin{equation}\label{eq:runratioabs}
\runratioabs = \frac{1}{\nsims} \left( 1+\sum_{i=2}^{\nsims} \bar{\delta} \left( |x_{i-1}|,|x_{i}| \right) \right).
\end{equation}
A further extension is to consider only the argument (i.e., phase) of the complex symbols,
\begin{equation}\label{eq:runratioarg}
\runratioarg =   \frac{1}{\nsims} \left( 1+\sum_{i=2}^{\nsims} \bar{\delta} \left( \measuredangle x_{i-1},  \measuredangle x_{i} \right) \right).
\end{equation}

\begin{example}[Run Ratios: Absolute Value and Argument]
Consider the QPSK sequences of Example~\ref{ex:runratios}. In this trivial example, \runratioabs is always $0.1$ because all symbols have the same amplitude, while \runratioarg is identical to the regular \runratio of \eqref{eq:runratio} because no two QPSK symbols share the same phase. This would obviously be different for higher-order QAM.
\end{example}

Note that the objective of the heuristic run ratios of Eqs.~\eqref{eq:runratio},~\eqref{eq:runratioabs}, and \eqref{eq:runratioarg} is not to exactly determine the absolute SNR variation after nonlinear fiber transmission, but to give some insight into whether the SNR variation with block length is in fact due to the temporal properties that is observed in a compound CC sequence. We will use these heuristics in the numerical analysis of the subsequent section to present strong arguments that it is indeed the induced structure when concatenating short CCDMs that leads to high effective SNR.

\section{Numerical Analysis}\label{sec:num_analysis}

\subsection{Simulation Setup}\label{ssec:sim_setup}
An idealized multi-span WDM fiber system is simulated to investigate the nonlinear fiber effects for varying block lengths. The simulation parameters are given in Table~\ref{tab:sim_params}. \rev{We focus on 64QAM and consider for illustration purposes only CC sequences with the fixed amplitude distribution $[0.4, 0.3, 0.2, 0.1]$, which was chosen as a coarse approximation to a Maxwell-Boltzmann distribution that allows to generate valid, i.e., integer-valued, composition for all block lengths that are multiplies of ten.} We note that the presented findings are expected to be similar for other high-order QAM formats and shaped distributions. The shaped symbols in both polarizations are generated from independent random data, and each WDM channel is modulated separately. When explicitly stated, an ideal interleaver that randomly shuffles the QAM symbols within each FEC block is included. The length per FEC word is 10800 64QAM symbols (64800 bits). After signal generation with root-raised cosine pulse shaping, the dual-polarization WDM signal is transmitted over 10 spans of standard single-mode fiber without polarization mode dispersion. Each span is followed by an Erbium-doped fiber amplifier (EDFA) with 6 dB noise figure. The propagation is simulated with the symmetrized split-step Fourier method with an adaptive step size such that the maximum nonlinear phase shift is $10^{-3}$ rad. At the receiver, the center channel is ideally filtered with a matched filter, chromatic dispersion is fully compensated, and any constant phase offset is removed in a genie-aided approach (see below). The effective SNR, averaged over both polarizations, is estimated from all pairs of the transmitted symbols $\Seqx$ (whose constellation has unit average energy) and received symbols $\Seqy$ as
\begin{equation}\label{eq:snr}
\text{SNR}_\text{eff} = 1/\text{var}(h\Seqy-\Seqx),
\end{equation}
where $\text{var}(\cdot)$ denotes variance and $h$ is a complex number that removes any constant phase offset by applying a single phase rotation and that scales the received symbols \rev{such that their conditional mean values are aligned with those of the transmit symbols, which is a similar approach to \cite[Sec.~III-B]{Buchali2016JLT_ProbShapingExp}.} The effective SNR includes all additive noise from the in-line amplifiers as well as nonlinear interference. A simulation run with approximately 500,000 QAM symbols is repeated ten times and the averaged effective SNR over all simulations is used as figure of merit in the following. We also show a 95\% confidence interval in the results via error bars.

\begin{table}[t]
\caption{Simulation parameters}
\centering
\renewcommand{\arraystretch}{1.1}
\begin{tabular}{c|c}
  \textbf{Parameter} & \textbf{Value} \\
  \hline
  Modulation & 64QAM \\
  Pol.-mux. & yes (independent pol's) \\
  Symbol rate & 32 GBd \\
  WDM spacing & 50 GHz \\
  WDM channels & 5 \\
  Pulse shape & root-raised cosine \\
  Roll-off & 10\% \\
  \hline
  Fiber length & 80 km \\
  No. of spans & 10 \\
  Fiber loss & 0.2 dB/km \\
  Dispersion & 17 ps/nm/km \\
  Nonlinearity $\gamma$ & 1.37 1/W/km \\
  Power per ch. & -0.5 dBm \\
  EDFA noise figure & 6 dB \\
  \hline
  Oversampling & 8$\times$ \\
  QAM symbols per run & $\sim$500,000 \\
  No. of simulation runs & 10 \\
  Amplitude PMF & $[0.4, 0.3, 0.2, 0.1]$ \\
  Block length $n$ & $10,\dots,10000$
  \end{tabular}
  \label{tab:sim_params}
\end{table}

\subsection{Results}\label{sec:num_results}

\subsubsection{Block Length Analysis}

\begin{figure}
\begin{center}
\inputtikz{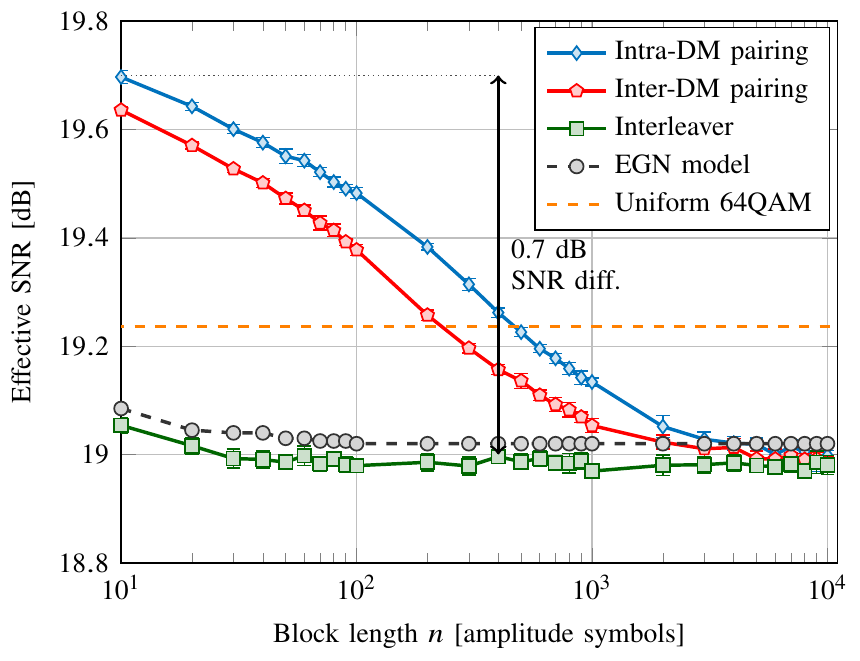}
\end{center}
\caption{Effective SNR after transmission in dB versus CCDM block length for 64QAM with one-sided amplitude distribution $[0.4, 0.3, 0.2, 0.1]$. \rev{Uniform 64QAM simulations and EGN model results for intra-DM pairing are included as references}.}
\label{fig:CCDM_SNR_vs_N}
\end{figure}

Figure~\ref{fig:CCDM_SNR_vs_N} shows the effective SNR as defined in \eqref{eq:snr} versus the CCDM block length $n$ for the above fiber system. We consider the two different amplitude-to-QAM pairing strategies introduced in the context of Fig.~\ref{fig:example_PAS_pairings} as well as interleaving within each FEC codeword. As shown previously in \cite{amariIntroducingEnumerativeSphere2019} for a different fiber setup, a clear dependence of SNR on the block length is observed when no interleaver is present and for the conventional intra-DM pairing (blue curve). The gap between ultra-short CCDM ($n=10$) and long CCDM ($n=10000$) amounts to 0.7 dB SNR. The effective SNR of shaped 64QAM with CCDM is approximately equal to that of uniform signaling at $n=500$. For longer block lengths, the SNR of shaped QAM is worse than for uniform, which is precisely what has been observed in the literature, see the introduction for a list of references. When the inter-DM pairing strategy is used (red curve in Fig.~\ref{fig:CCDM_KLdiv_vs_N}), SNR is also inversely proportional to $n$, but with approximately 0.06~dB offset compared to intra-DM pairing up to a block length of about 1000. We believe that the reason for this difference is that the inherent shuffling discussed in Sec.~\ref{ssec:diff_sequences} is not as distinct when the QAM symbols are combined from two independent DMs, as it is done for inter-DM pairing. When an interleaver is included (green curve), the block-length dependence of the SNR has completely disappeared except for a very small difference of 0.05~dB SNR at very short lengths where the distribution is different from the one at larger $n$, see Fig.~\ref{fig:CCDM_KLdiv_vs_N}. This means that the SNR decrease with increasing $n$ reported in Fig.~\ref{fig:CCDM_SNR_vs_N} is due to some nonlinear temporal interactions that occur within each DM block and also between adjacent blocks.

\rev{Further included in Fig.~\ref{fig:CCDM_SNR_vs_N} are the results of the EGN model \cite{Carena2014OptExp_EGNmodel}, evaluated with 100,000 Monte Carlo samples and including self- and cross-channel interference, for the intra-DM pairing strategy, for which the kurtosis was analyzed in Fig.~\ref{fig:CCDM_Kurtosis_vs_N}. As expected, the EGN curve is mostly flat as the EGN model is by design unable to predict time-dependent NLI effects.}

We stress that we only show effective SNR in Fig.~\ref{fig:CCDM_SNR_vs_N}, which does not take into account the DM rate loss. The best operating point \rev{in terms of maximizing the achievable information rate (AIR) for finite-length CCDM} is not at ultra-short lengths because the CCDM rate loss is prohibitively large in this regime, see, e.g., \cite[Fig.~6]{Fehenberger2019TCOM_MPDM}. \rev{In further simulations (not shown herein) we find that the optimal block length for the considered system is at approximately 500~symbols, with longer block lengths having slightly lower AIR, which is in good agreement with \cite[Fig.~7]{amariIntroducingEnumerativeSphere2019}. }

When we recall that it was argued in Sec.~\ref{ssec:diff_sequences} that few long DMs can generate the same sequence as many short DMs, the question might arise why the SNR in Fig.~\ref{fig:CCDM_SNR_vs_N} is worse for large $n$ than for small $n$. A possible answer is that the probability of a particular CCDM sequence being generated is $\mathcal{A}^{-n}$ where $\mathcal{A}$ is the number of shaped amplitudes, which is 4 in the case of 64QAM. Even for a moderate length of $n=200$, this number is $3.88\cdot 10^{-121}$, which shows how unlikely it is that such a CCDM happens to output a sequence that has the same temporal properties (e.g., number of runs) as an ultra-short CCDM.


The implications of these findings are twofold. Firstly, if an interleaver is included in the system, all blocks lengths perform approximately the same; there is no block-length-dependent SNR but only an approximately constant one. Secondly, as outlined in Sec.~\ref{ssec:diff_sequences}, any group of short-length CCDM blocks can, in principle, be generated by a single long CCDM. Therefore, it should be possible for a fixed large $n$ to modify which CCDM sequences are selected such that an SNR gain is achieved. It is an open research problem to find algorithms that enable such a nonlinearity-aware mapping and also to identify a priori which sequences are good and which ones are bad. \rev{Note that this can only be realized without additional rate loss when the DM mapping function is not bijective, i.e., when there are output sequences that cannot be reached regardless of the binary input word.} As indicated in Fig.~\ref{fig:CCDM_KLdiv_vs_N} by the SNR results for interleaving, no structure, i.e., random shuffling, seems not to be the way forward. This means that some structure in the output sequence (as it is introduced by concatenating many short CCDMs) is beneficial for increasing the SNR after fiber transmission. A first demonstration of nonlinearity-tailored distribution matching is presented in \cite{Fehenberger2019ECOC_NLI_MPDM} where the construction of an MPDM is modified to increase the SNR of the worst shaping block as well as the average SNR.

\subsubsection{Run Ratio Analysis}
\begin{figure}
\begin{center}
\inputtikz{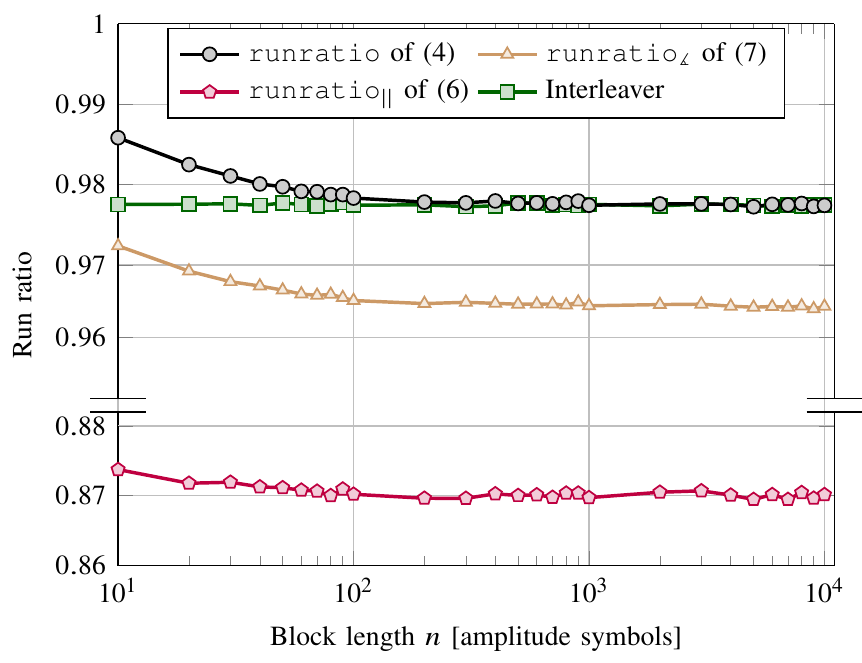}
\end{center}
\caption{Run ratios versus block length $n$. Only intra-DM pairing is considered.}
\label{fig:CCDM_SNR_vs_RunLength}
\end{figure}

In the following, we use the run ratios defined in \eqref{eq:runratio}, \eqref{eq:runratioabs}, and \eqref{eq:runratioarg} as proxies to study the structure within the DM codewords. Figure~\ref{fig:CCDM_SNR_vs_RunLength} shows, for the same simulation data that was used in Fig.~\ref{fig:CCDM_SNR_vs_N}, run ratio as a function of block length. Focusing on the regular run ratio \runratio of \eqref{eq:runratio} first, we observe that short blocks (which resulted in high SNR) indeed have a higher run ratio. We recall that for a fixed sequence length, an increase in run ratio means a larger number of runs and thus, some shuffling in the sequence. In combination with the results presented in Fig.~\ref{fig:CCDM_SNR_vs_N}, this suggests for the considered setup that when long runs of identical symbols are prohibited, or equivalently when the concentration of symbols is limited, a high effective SNR is achieved. Note that Fig.~\ref{fig:CCDM_SNR_vs_RunLength} shows a block-length-dependent run ratio only up to approximately $n=100$, \rev{and simulations with the average run length as figure of merit found qualitatively the same trend as \runratio.} This indicates that more refined metrics are required to fully capture the time-dependent nonlinear interactions of longer blocks. For example, any metric that is based only on the number of runs does not differentiate between whether a high- or low-energy symbol constitutes a run, which could, however, have a significant impact on the magnitude of the induced nonlinear interactions. Using the notation of Example~\ref{ex:runratios}, the two sequences $[\alpha \alpha \alpha \alpha \beta \gamma \beta \gamma \delta \beta ]$ and $[\alpha \alpha \beta \alpha \alpha \beta \beta \gamma \delta \gamma ]$ have the same run ratio of 0.7, but there is no reason to assume that they induce an identical amount of nonlinear interference. \rev{Furthermore, we note that when the entire QAM sequences is sorted prior to transmission, the run ratio does not give a clear relationship with effective SNR.}\footnote{\rev{Albeit theoretically interesting, we consider this case somewhat unrealistic since by sorting every sequence, the amount of information that is conveyed over the channel becomes vanishingly small.}}

When an interleaver is inserted into the system, the run ratio is independent of the block length, as expected from the approximately flat SNR-curve of Fig.~\ref{fig:CCDM_SNR_vs_N}. The run ratio for  inter-DM pairing (not shown in Fig.~\ref{fig:CCDM_SNR_vs_RunLength}) follows roughly the same behavior as intra-DM pairing, indicating that the heuristic run ratio is able to qualitatively describe the trend of SNR dependence, but is unable to predict the exact SNR value. Further included in Fig.~\ref{fig:CCDM_SNR_vs_RunLength} for intra-DM pairing are the absolute-value and phase run ratios defined in \eqref{eq:runratioabs} and \eqref{eq:runratioarg}, respectively. We observe that \runratioabs has very little variation for the considered block lengths, whereas \runratioarg follows the same behavior as \runratio. This could suggest that amplitude jumps are not the main reason for the SNR variation with block length, but phase variations have a stronger impact. A thorough analysis of this aspect is part of our ongoing work. 

\section{Conclusions}
We have investigated constant-composition sequences of varying lengths and have confirmed in extensive simulations of a multi-span WDM system that the effective SNR after transmission is indeed decreasing with increasing block length, as previously noted in \cite{amariIntroducingEnumerativeSphere2019}. The reasons for this effect are twofold. Firstly, for very short blocks, combining two shaped amplitudes into the shaped QAM payload, as it is typically done in the PAS framework, can lead to certain high-energy symbols occurring more rarely than for longer blocks. Hence, varying the CCDM block length for a fixed amplitude distribution can impact the QAM distribution, which in turn impacts the fiber nonlinearities and thus the SNR. Secondly, and more importantly, we have studied properties of long compound sequences that consists of many short CCDM blocks, or just a few long blocks. We explain that a key characteristic of short sequences is an inherent shuffled structure which prohibits long runs of identical symbols. To further study this temporal property, a heuristic called run ratio is introduced which describes the number of consecutive-symbol runs in a sequence. A numerical study of this metric for compound sequences of short-to-moderate length suggests that such a limited symbol concentration is beneficial as weaker nonlinear fiber effects occur. We believe that this finding could motivate and facilitate the search for signaling schemes that suppress the generation of nonlinear interference.

\section*{Acknowledgments}
T. Fehenberger would like to thank Dr. Alex Alvarado (TU Eindhoven) for inspiring discussions on the problem at hand as well as Kaiquan Wu (TU Eindhoven) for valuable comments on the first version of this manuscript.


\end{document}